\documentclass[10pt,conference]{IEEEtran}
\usepackage{epsfig}
\usepackage{amssymb,amsmath}
\usepackage{graphicx}
\usepackage{epic}
\usepackage{color}
\newcommand {\bE} {\mathbb{E}}
\newcommand{\openone}{\leavevmode\hbox{\small1\normalsize\kern-.33em1}}

\newcommand {\beq} {\begin{equation}}
\newcommand {\eeq} {\end{equation}}

\newtheorem{theorem}{Theorem}{}
\newtheorem{lemma}{Lemma}{}
{}
{}
{}
{}

\newcommand {\noise} {l^n}

\newcommand {\code} {{\mathcal C}}

\newcommand {\dcode} {{\mathcal C}^{\perp}}

\newcommand {\dlm} {\texttt{l}_{\text{max}}}
\newcommand {\drm} {\texttt{r}_{\text{max}}}

\begin{document}

\title{Decay of Correlations in Low Density Parity Check Codes: Low Noise Regime}
\author{\authorblockN{Shrinivas Kudekar and Nicolas Macris}
\authorblockA{Laboratory of theory of communications \\ 
School of Computer and Communication Sciences, EPFL, Lausanne.\\
Email: {\ttfamily \{shrinivas.kudekar,nicolas.macris\}@epfl.ch}}
}

\maketitle

\begin{abstract} Consider transmission over a binary additive white gaussian
noise channel using a fixed low-density parity
check code. We consider the posterior measure over the code bits and the corresponding correlation between two codebits, averaged over the noise realizations. We show that for low enough noise variance this average correlation decays
exponentially fast with the graph distance between the code bits. One consequence of this result is that for low enough noise variance the GEXIT functions (further averaged over a standard code ensemble) of the belief propagation and optimal decoders are the same.
\end{abstract}

\section{Introduction}\label{section1}

We consider transmission over a binary additive white gaussian
noise channel (BIAWGN) using low-density parity
check codes (LDPC) and the optimal MAP decoder. We are interested in the behavior of the correlation between two code bits as a function of their graph distance. In \cite{KuMa07TC} we treated this problem, for the regime of high noise, for a special code ensemble containing a sufficiently large fraction of degree one variable nodes. In the present contribution we attack the problem {\it in the low noise regime}.

The behavior of correlations between relevant degrees of freedom is of central interest in the analysis of Gibbs measures, and various approaches have been developed to tackle such problems. The Gibbs measures associated with the optimal decoder of LDPC codes confront us with new challenges which invalidate the direct use of the standard methods. For example it is easy to see that the standard Dobrushin type methods \cite{Georgii} fail due to the presence of hard constraints. In the high noise regime we were able to convert the problem (at least in the special case of \cite{KuMa07TC}) to a spin glass containing a mixture of soft and hard constaints for which appropriate cluster expansions can be applied. These expansions have been applied to the simpler case of low density generator matrix codes (LDGM) in \cite{KuMa08itw} for the high noise regime, which boils down to a high temperature spin glass.

The low noise regime which is our interest here is a truly low temperature spin glass problem for which all the above methods fail. The general idea of our strategy is to apply a duality transformation to the LDPC Gibbs measure. It turns out that the dual problem does not correspond to a well defined communications problem, and in fact it does not even correspond to a well defined Gibbs measure because the ``weight'' takes positive as well as negative values. Nevertheless the dual problem has the flavor of a high noise LDGM system (or high temperature spin glass) and we are able to treat it through cluster expansions. There exist a host of such expansions \cite{Brydges}, but we wish to stress that the simplest ones do not apply to the present situation for at least two reasons. The first, is that there exist arbitrarily large portions of the dual system which are in a low noise (or low temperature) phase with positive probability (this is related to the Griffith singularity phenomenon \cite{Frohlich}). The second, is that the weights of the dual problem are not positive so that the method in \cite{KuMa08itw} does not work. It turns out that a cluster expansion originaly devised by Berretti \cite{Berretti} is very well suited to overcome all these problems. 

Our analysis can also be carried through for a class of other channels including the BSC and BEC, but we do not give the details here. The case of the BEC is special because under duality the Gibbs weight remains positive and  the communication problem using LDPC codes on ${\rm BEC}(\epsilon)$ transforms to a real communication problem using LDGM codes on the ${\rm BEC}(1-\epsilon)$ \cite{BrinkAshikmin}.

In the last section we sktech an application of our main result to the MAP-GEXIT function (in other words the first derivative of the input-output entropy with respect to the noise parameter). We prove that in the low noise regime where the average correlation decays (fast enough) the MAP-GEXIT function can be exactly computed from the density evolution analysis. These curves remain non-trivial all the way down to zero noise as long as there are degree one variable nodes (e.g Poisson LDPC codes). This proves that a non-trivial replica solution is the exact expression for the input-output entropy of a class of LDPC codes (containing a fraction of degree one variable nodes) on the BIAWGN channel. 
Previously the replica expression was only known to be a one-sided bound \cite{Montanari}, \cite{KuMa06isit} for general ensembles and channels. The equality had been obtained previously for some ensembles on the BEC using duality \cite{CyRuMon07isit} and the interpolation method \cite{KuKoMa}.

\section{Decay of Correlations}\label{section2}

Let  $x^n$ be a binary codeword of length $n$ from a fixed LDPC code with bounded, but otherwise arbitrary variable and check node degrees. In the sequel we call $\dlm, \drm$ the maximal variable and check degrees.
The noise variance of the
BIAWGN channel is $\epsilon^2$ and $y^n$ denotes the received message. Assuming without loss of generality that the channel input is the all zero codeword, the output can be mapped onto the half-log-likelihood ratio $l_i = \frac12\ln\frac{p_{Y\vert X}(y_i\vert 1)}{p_{Y\vert X}(y_i\vert 0)}$ where $p_{Y\vert X}(y\vert x)$ is the channel's transition matrix. The channel outputs are i.i.d with distribution $p_{Y\vert X}(y\vert 0)$ which induces a distribution $c(l)=\frac{1}{\sqrt{2 \pi
\epsilon^{-2}}}e^{-(l-\epsilon^{-2})^2/2\epsilon^{-2}}$. Mapping the codewords $x^n$ to spin configurations $\sigma^n$ with $\sigma_i=(-1)^{x_i}$, the posterior measure becomes (for a uniform prior)
\begin{equation}\nonumber
p_{X^n\vert Y^n}(x^n\vert y^n)=\frac{1}{Z_P}\prod_{i=1}^n e^{l_i\sigma_i} \prod_{c=1}^m \frac12 (1+\prod_{i\in c}\sigma_{i})
\end{equation} 
In this expression $\prod_{c}$ is a product over all the parity check constraints of the code and $\prod_{i\in c}$ is a product over variable
nodes attached to the check node $c$. The partition function $Z_P$ is simply the normalizing factor
\begin{equation}\nonumber
Z_P=\sum_{\sigma^n\in\{-1,+1\}^n}
\prod_{i=1}^n e^{l_i\sigma_i} \prod_{c=1}^m \frac12 (1+\prod_{i\in c}\sigma_{i})
\end{equation} 
The average of an arbitrary function $f(\sigma^n)$ with respect to the above measure is denoted as  $\langle f(\sigma^n)\rangle_P$ where the subscript $P$ refers to parity check (later we use various other brackets). This is still a random quantity which depends on the channel output realization. Further averages with respect to the noise are denoted by 
$\mathbb{E}_{l^n}[\langle f(\sigma^n)\rangle_P]$. Of course it does not make sense to permute the expectation $\mathbb{E}_{l^n}$ and the bracket $\langle - \rangle_P$ because of the normalizing factor $Z_P$ in the denominator.

Our main result is on the average correlation between any two codebits defined by 
\begin{equation}\nonumber
C_P(i,j)=\mathbb{E}_{l^n}[\vert\langle \sigma_i\sigma_j \rangle_P - \langle \sigma_i\rangle_P \langle \sigma_j\rangle_P\vert]
\end{equation}
\begin{theorem}[Decay of Correlations]\label{theorem1}
Consider transmission over a BIAWGN channel with noise variance $\epsilon^2$ using an arbitrary fixed LDPC code. Set $k=(\dlm\drm)^{1/2}$. Let $\text{dist}(i,j)$ denote the graph distance between the codebits $i,j$. There exist strictly positive purely numerical constants $\epsilon_0$, $c_1$, $c_2$ such that for $\epsilon^2\leq \epsilon_0^2 k^{-2}(\ln k)^{-1}$ and $\text{dist}(i,j)>4\dlm$ we have
\begin{equation}\label{firstdecay}
C_P(i,j)\le c_1 e^{-\frac{c_2}{\epsilon^2 k}\text{dist}(i,j)}
\end{equation}
\end{theorem}
\vskip 0.25cm
\noindent{\it Remark}: By graph distance we mean the smallest possible number of edges on a path connecting $i$ and $j$.

In fact we will derive (and use in section \ref{section5}) a slightly more general estimate. Suppose that the bits $x_i$ are transmitted at different noise levels 
$\epsilon_i\leq \epsilon$. Then 
\begin{equation}\label{general}
C_P(i,j)\le c_1 e^{-c(\frac{1}{\epsilon_i^2}+\frac{1}{\epsilon_j^2})}e^{-\frac{c_2}{\epsilon^2 k}\text{dist}(i,j)}
\end{equation} 
where $c>0$ is a strictly positive number. In particular if bits $x_i$ or $x_j$ are perfectly received we recover $C_P(i,j)=0$.

\section{Duality Formulas}\label{section3} 

A general theory of duality for codes on graphs can be found in \cite{Forney} and references therein. Here we derive by elementary means formulas that are useful to us.
Let $\code$ be a binary parity check code and $\dcode$ its dual. We apply the Poisson summation formula 
\begin{equation}\nonumber
\sum_{\sigma^n \in\code} f(\sigma^n) = \frac{1}{|\dcode|}
\sum_{\tau^n\in \dcode} \widehat{f}(\tau^n)
\end{equation}
where the Fourier (or Hadamard) transform is,
\begin{equation}\nonumber
\widehat{f}(\tau^n) = \sum_{\sigma^n \in \{-1,+1\}^n} f(\sigma^n) e^{i\frac{\pi}{4} \sum_{j=1}^n (1-\tau_j)(1-\sigma_j)}
\end{equation}
to the partition function $Z_P$ of an LDPC code $\code$. The dual code $\dcode$ is an LDGM with codewords given by
$\tau^n$ where 
\begin{equation}\label{codewords}
\tau_i=\prod_{c\in i} u_c
\end{equation}
and $u_c$ are the $m$ information bits ($i$ and $c$ will always refer to the variable and check nodes of the original LDPC Tanner graph and $c\in i$ means that $c$ is connected to $i$). A straigthforward application of the Poisson formula then yields an extended form of the MacWilliams identity,
\begin{equation}\label{duality}
 Z_P= \frac{1}{\vert\dcode\vert}e^{\sum_{j=1}^n l_j}Z_G
\end{equation}
where 
\begin{equation}\nonumber
Z_G  =  \sum_{u^m\in\{-1,+1\}^m} \prod_{i=1}^n (1+ e^{-2l_i}\prod_{c\in i} u_c)
\end{equation}
This expression formaly looks like the partition function of an LDGM code (hence the subscript G) with ``channel log-likelihoods'' $g_i$ such that $\tanh g_i = e^{-2l_i}$. This is truly the case for the BEC($\epsilon$) where $l_i=0,+\infty$ and hence $g_i=+\infty,0$ which still correspond to a BEC($1-\epsilon$). The logarithm of partition functions is related to the input-output entropy and one recovers (taking the $\epsilon$ derivative) the well known duality relation between EXIT functions of a code and its dual on the BEC \cite{BrinkAshikmin}. For other channels however this is at best a formal (but useful) analogy since the weights can be negative or equivalently the $g_i$ can assume complex values.

We will need a duality formula for the correlations themselves. We introduce a bracket $\langle-\rangle_G$ which {\it is not} a true probabilistic expectation (but it is linear)
\begin{equation}\nonumber
\langle f(u^m)\rangle_G = \frac{1}{Z_G}
\sum_{u^m\in\{-1,+1\}^m} f(u^m)\prod_{i=1}^n (1+ e^{-2l_i}\prod_{c\in i} u_c)
\end{equation}
The denominator may vanish, but it can be shown that when this happens the numerator also does so, and in a way that ensures the finiteness of the ratio (this will become quite clear in all our subsequent calculations).
Taking logarithm of \eqref{duality} and then the derivative with respect to $l_i$ we find
\begin{align}\label{firstderivative}
\langle \sigma_i \rangle_P
& = \frac{1}{\tanh 2l_i} -\frac{\langle\tau_i\rangle_G}{\sinh 2l_i} 
\end{align}
and differentiating once more with respect to  $l_j$, $j\neq i$ 
\begin{align}\label{secondderivative}
\langle \sigma_i \sigma_j\rangle_P - \langle \sigma_i\rangle_P\langle \sigma_j\rangle_P  =  \frac{\langle \tau_i\tau_j \rangle_G - \langle \tau_i \rangle_G\langle \tau_j \rangle_G}{\sinh 2l_i\sinh 2l_j}
\end{align}
We stress that in \eqref{firstderivative}, \eqref{secondderivative}, $\tau_i$ and $\tau_j$ are given by products of information bits \eqref{codewords}. The left hand side of \eqref{firstderivative} is obviously bounded. It is less obvious to see this directly on the right hand side and here we just note that the pole at $l_i=0$ is harmless since, for $l_i=0$, the bracket has all its ``weight`` on configurations with $\tau_i=1$. Similar remarks apply to \eqref{secondderivative}.
In any case, we will beat the poles by  using the following trick. For any $0<s<1$ and $\vert x\vert\leq 1$ we have $\vert x\vert\leq \vert x\vert^s$, thus
\begin{equation}\nonumber
  C_P(i,j) \leq 2^{1-s} \mathbb{E}_{l^n}[\vert
\langle\sigma_i\sigma_j\rangle_P - \langle\sigma_i\rangle_P \langle\sigma_j\rangle_P\vert^s]
\end{equation}
and using \eqref{secondderivative} and Cauchy-Schwarz
\begin{align}\label{spower}
 C_P(i,j) & \leq 2^{1-s}\mathbb{E}[(\sinh 2l)^{-2s}]
\\ \nonumber &
\times
\mathbb{E}_{l^n}[\vert \langle \tau_i\tau_j \rangle_G - \langle \tau_i \rangle_G\langle \tau_j \rangle_G\vert^{2s}]^{1/2}
\end{align}
The following bound
\begin{equation}\label{sinhbound}
\mathbb{E}[(\sinh 2l)^{-2s}]
\leq
\frac{c}{\vert 1-2s\vert}e^{-c^\prime\frac{s(1-2s)}{\epsilon^{2}}}
\end{equation}
on the prefactor turns out to be important in our analysis. Here $0<s<\frac{1}{2}$ and $c>0$, $c^\prime>0$ are purely numerical constants.

\section{Proof of Main Theorem}\label{section4}

From inequalities \eqref{spower}, \eqref{sinhbound} of the previous section we see that it suffices to prove that 
\begin{equation}\nonumber
C_G(i,j;s) = \mathbb{E}_{l^n}[\vert \langle \tau_i\tau_j \rangle_G - \langle \tau_i \rangle_G\langle \tau_j \rangle_G\vert^{2s}]
\end{equation}
decays.
As explained in the introduction, the main tool used here is a cluster expansion of Berretti \cite{Berretti} (that has the advantage of dealing simultaneously with the Griffith singularity phenomenon and at the same time does not use the positivity of the weights).
Here we can only explain the resulting expansion, adapted to our setting, without giving the full derivation (a good starting point is \cite{Frohlich}). We have
\begin{equation}\nonumber
\langle \tau_i\tau_j\rangle_G - \langle \tau_i\rangle_G\langle \tau_j\rangle_G 
= \frac12 \sum_{\hat{X}} K_{i,j}(\hat{X}) \Bigl(\frac{Z_{G}(\hat{X}^c)}{Z_G}\Bigr)^2 
\end{equation}
where 
\begin{equation}\nonumber
K_{i,j}(\hat{X}) \triangleq \sum_{\substack{u_c^{(1)}, u^{(2)}_c \\ c
\in \hat{X} }}\sum_{\substack{\Gamma ~\text{compatible} \\
\text{with} \hat{X}}} 
(\tau_i^{(1)}-\tau_i^{(2)})(\tau_j^{(1)}-\tau_j^{(2)}) \prod_{k\in \Gamma} E_k 
\end{equation}
and 
\begin{align}\label{eq:Ea}
E_k = \tau_k^{(1)} e^{-2 l_k} +\tau_k^{(2)} e^{-2 l_k} +\tau_k^{(1)}\tau_k^{(2)} e^{-4 l_k}
\end{align}
Here $u_c^{(1)}$ and $u_c^{(2)}$ are two independent copies of the information bits (these are also known as real
replicas) and $\tau_k^{(\alpha)}= \prod_{c\in k} u_k^{(\alpha)}$. To explain what are $\hat{X}$ and $\Gamma$ we keep referring to checks and variables in the original LDPC Tanner graph language: checks are indexed by $c$ and variables by $i$. Given a subset $S$ of variable or ckeck nodes of the Tanner graph let $\partial S$ be the subset of neighboring nodes.
The sum over $\hat{X}$ is carried over clusters of check nodes such that: (i) $\hat{X}$ is ''connected via hyperedges'' (this means that $\hat{X}=\partial X$ for some connected subset $X$ of variable nodes; $X$ is connected if any pair of  variable nodes can be joined by a path all of whose variable nodes lie in $X$)  and (ii) $\hat{X}$ contains both the $\partial i$ and
$\partial j$. $\Gamma$ is a set of variable nodes (all distinct). We say that $\Gamma$ 
is compatible with $\hat{X}$ if: (i) $\partial\Gamma\cup
\partial i\cup\partial j = \hat{X}$, (ii) $\partial\Gamma \cap
\partial i \neq \phi$ and $\partial\Gamma \cap
\partial j \neq \phi$, (iii) there is a walk connecting $\partial i$ and $\partial j$ such that all its variable nodes are in $\Gamma$.
Finaly,
\begin{equation}\nonumber
Z_{G}(\hat{X}^c)  = \sum_{\substack{u_c \\ c\in  \hat{X}^c}} 
\prod_{\substack{\text{all}\; i \;\text{s.t.} \\ \partial i \cap \hat{X} =
\phi}} (1 + e^{-2 l_i}\prod_{c\in i} u_c )
\end{equation}

Using
$\vert\sum_{i}a_i\vert^{2s} \le \sum_{i} \vert a_i\vert^{2s}
$ for $0<2s<1$ and then Cauchy-Schwarz,
we find 
\begin{equation}\nonumber
C_G(i,j;s) \leq 
\frac{1}{2}\sum_{\hat{X}} T_1(\hat X) T_2(\hat X) 
\end{equation}
where
\begin{equation}\label{T1}
T_1(\hat X)^2 = \bE_{\noise}\bigl[|K_{i,j}(\hat{X})|^{4s}\bigr] 
\end{equation}
and 
\begin{equation}\label{T2}
T_2(\hat X)^2 =\bE_{\noise}\bigl[\Big(\frac{Z_{G}(\hat{X}^c)}{Z_G}\Big)^{8s}\bigr] 
\end{equation}
\noindent{\it Bound on $T_1(\hat X)$}. Trivially bounding the spins in $\eqref{eq:Ea}$ by $1$ we deduce (in the first inequality we need $4s<1$ and in the second $8s<1$)
\begin{align}\nonumber
T_1(\hat X)^2 & \leq 4^{\vert \hat X\vert}\sum_{\substack{\Gamma ~\text{compatible} \\
\text{with} \hat{X}}}(2^{4s}\bE_{\noise}[e^{-8sl}] + \bE_{\noise}[e^{-16sl}])^{\vert \Gamma\vert}
\\ \nonumber &
\leq
 4^{\vert \hat X\vert}
\sum_{\substack{\Gamma ~\text{compatible} \\
\text{with} \hat{X}}} 2^{(4s+1)\vert\Gamma\vert}
e^{-\frac{8s(1-8s)}{\epsilon^2}\vert \Gamma\vert}
\end{align}
Now let us set $s=\frac{1}{16}$ and take 
$\epsilon^2\leq (10\ln 2)^{-1}$ for simplicity. The bound becomes
\begin{equation}\nonumber
T_1(\hat{X})^2 \leq  4^{\vert \hat X\vert}
\sum_{\substack{\Gamma ~\text{compatible} \\
\text{with} \hat{X}}} 
e^{-\frac{1}{8\epsilon^2}\vert \Gamma\vert}
\end{equation}
If $\Gamma$ is compatible with $\hat X$ we necessarily have
$\vert \partial\Gamma\vert \geq |\hat{X}| -
\vert\partial i\vert - \vert\partial j\vert$ an since $\vert\partial \Gamma\vert \leq \vert \Gamma\vert \dlm$, we get $\vert \Gamma\vert \geq 
(|\hat{X}| -2 \dlm)/\dlm$. Also, the maximum
number of variable nodes which have an intersection with $\hat{X}$ is
$|\hat{X}|\drm$. Thus there are at most $2^{|\hat{X}|\drm}$ possible choices for
$\Gamma$. These remarks imply
\begin{equation}\nonumber
T_1(\hat X)^2  \le  2^{(2+\drm)|\hat{X}|} e^{-\frac{1}{8\epsilon^2}(|\hat{X}| -2 \dlm)/\dlm}
\end{equation}

\noindent{\it Bound on $T_2(\hat X)$}. The ratio 
\eqref{T2} is not easily estimated directly because 
the weights in $Z_G$ are not positive. However we can use the duality transformation \eqref{duality} to get a new ratio of partition functions with positive weights,
\begin{equation}\nonumber
\frac{Z_{G}(\hat{X}^c)}{Z_G} = \biggl(\exp{\sum_{\substack{\text{all}\,i\, \text{s.t} \\ \partial i \cap \hat{X} \neq\phi}} l_i}\biggr)\frac{\vert\dcode(\hat{X}^c)\vert}{\vert\dcode\vert} \frac{Z_{P}(\hat{X}^c)}{Z_P}
\end{equation}
with
\begin{equation}\nonumber
Z_{P}(\hat{X}^c) = \sum_{\substack{\sigma_i \\
\partial i \cap \hat{X} =\phi}} \prod_{\substack{\text{all}\,i\, \text{s.t} \\ \partial i \cap \hat{X} =\phi}} e^{l_i\sigma_i} \prod_{c\in \hat{X}^c} \frac{1}{2}(1+\prod_{\substack{i\in c\,\text{and}\\ \partial i\cap\hat{X}=\phi}}\sigma_i)
\end{equation}
which is the partition function corresponding to the subgraph (of the full Tanner graph) induced by checks of $\hat{X}^c$ and variable nodes $i\,\text{s.t}\,\partial i\cap\hat{X}=\phi$. Moreover $\dcode(\hat{X}^c)$ is the dual of the later code $\code(\hat{X}^c)$ defined on the subgraph. By standard properties of the rank of a matrix, the rank of the parity check matrix of $\code(\hat{X}^c)$, which is obtained by removing rows (checks) and columns (variables) from the parity check matrix of $\code$, is smaller than the rank of the parity check matrix of $\code$. Thus $\vert\code(\hat{X}^c)\vert \geq \vert\code\vert$ and $\vert\dcode(\hat{X}^c)\vert \leq \vert\dcode\vert$. Moreover 
\begin{equation}\nonumber
\biggl(\exp{\sum_{\substack{\text{all}\,i\, \text{s.t} \\ \partial i \cap \hat{X} \neq\phi}} l_i}\biggr)
Z_{P}(\hat{X}^c) \leq Z_P
\end{equation}
To see this one must recognize that the left hand side is the sum of terms of $Z_P$ corresponding to $\sigma^n$ such that $\sigma_i=+1$ for $\partial i\cap\hat{X}\neq \phi$ (and all terms are $\geq 0$). These remarks imply for \eqref{T2}
\begin{equation}\nonumber
T_2(\hat{X})^2\leq 1
\end{equation}

Now we can conclude the proof of theorem \ref{theorem1}. From the bounds on \eqref{T1} and \eqref{T2} we get for 
$\epsilon^2<(\dlm(2+\drm)16\ln 2)^{-1}$ 
\begin{equation}\nonumber
C_G(i,j;s=\frac{1}{16})\leq \sum_{\hat{X}}  e^{-\frac{1}{32\dlm\epsilon^2}(\vert\hat{X}\vert -2\dlm)}
\end{equation}
The clusters $\hat{X}$ connect $\partial i$ and $\partial j$ and thus have sizes 
$\vert\hat{X}\vert\geq \frac{1}{2}\text{dist}(i,j)$. Moreover the number of clusters of a given size grows at most like $(\dlm\drm)^{\vert \hat{X}\vert}$. Working out the final bounds, and putting them in a symmetrical form, the net result is that for $\text{dist}(i,j)> 4\dlm$ we can find a purely numerical constant $\epsilon_0$ such that for $\epsilon^2<\epsilon_0^2 k^{-2}(\ln k)^{-1}$ 
\begin{equation}\nonumber
C_G(i,j;s=\frac{1}{16})\leq 
c_1 e^{-\frac{c_2}{\epsilon^2 k}\text{dist}(i,j)}
\end{equation}
where $k= (\dlm\drm)^{\frac{1}{2}}$ and $c_1$ and $c_2$ a strictly positive numbers.
Using this bound with \eqref{spower} and \eqref{sinhbound}  concludes the proof of \eqref{firstdecay} and \eqref{general}.

\section{Exactness of Density Evolution}\label{section5}

In this section we illustrate an application of the theorem
to the GEXIT function of standard irregular LDPC ensembles with degrees bounded by $\dlm,\drm$. Let $h_n=\frac{1}{n}H(X^n\vert Y^n)$ be the input-output entropy. The MAP-GEXIT function is in general  defined as 
\begin{equation}\nonumber
\frac{d}{d(\epsilon^{-2})}\mathbb{E}_{\text{LDPC}}[h_n]
\end{equation}
\begin{theorem}[Exactness of Density Evolution]\label{theorem2}
One can find a strictly positive number $\epsilon_1$ (in general smaller than the $\epsilon_0$ of theorem \ref{theorem1}) such that for 
$\epsilon^2\leq \epsilon_1^2k^{-2}(\ln k)^{-1}$ 
\begin{align}\nonumber
\lim_{n\to \infty}\frac{d}{d\epsilon^{-2}} 
\bE_{\text{LDPC}} [h_n] = \frac{1}{2}(\lim_{d\to \infty}\bE_{\text{LDPC},l}[\tanh(l+\Delta^{(d)})] -1)
\end{align}
where $\Delta^{(d)}$ is the soft bit-estimate given by the density evolution analysis of the BP decoder.
\end{theorem}
\vskip 0.25cm
The proof of this theorem rests on the simple formula \cite{MacTransactions07}, \cite{MacTransactions072} valid for the BIAWGN channel
\begin{equation}\label{eq:awgnfderi}
 \frac{d}{d(\epsilon^{-2})} \bE_{\code} [h_n] = \frac{1}{2}
(\mathbb{E}_{\text{LDPC},\noise}[\langle \sigma_o\rangle_{P}]-1)
\end{equation} 
where the variable node $o$ is selected uniformly at random (the result is independent of the node due to symmetry). In this formula $\mathbb{E}_{\noise}[\langle\sigma_o\rangle]$ is the MAP soft-bit estimate. 

In fact one can verify that the density evolution analysis is equivalent to performing statistical mechanical sums on 
a tree whose leaves are the spins (variable nodes) with free boundary conditions (channel outputs as initial conditions). More precisely if we call $N_d(o)$ the neighborhood of depth $d$ of $o$ for $d$ even (that is all the nodes of the Tanner graph that are at a distance $\leq d$ from $o$) and consider the LDPC Gibbs measure $\langle -\rangle_{N_d(o)}$ restricted to the subgraph $N_d(o)$, we can verify by explicit calculation that
\begin{equation}\nonumber
\bE_{\text{LDPC},l}[\tanh(l+\Delta^{(d)})] = \mathbb{E}_{\text{LDPC},\noise}[\langle \sigma_o\rangle_{N_d(o)}\vert N_d(o)~\text{is a tree}]
\end{equation}
Now for $d$ fixed, $N_d(o)$ is a tree with probability $1-O(\frac{\gamma^d}{n})$ where $\gamma$ depends only on the maximum node degrees, so
\begin{equation}\label{tan}
\mathbb{E}_{\text{LDPC},\noise}[\langle \sigma_o\rangle_{N_d(o)}] = \bE_{\text{LDPC},l}[\tanh(l+\Delta^{(d)})] +O(\frac{\gamma^d}{n}) 
\end{equation}
Thus in view of \eqref{eq:awgnfderi} the theorem will follow if we can show that
\begin{equation}\label{xi}
\mathbb{E}_{\noise}[\langle \sigma_o\rangle_{P}]=\mathbb{E}_{\noise}[\langle \sigma_o\rangle_{N_d(o)}] + O(e^{-\xi\frac{d}{\epsilon^2}})
\end{equation}
with $\xi>0$ and $O(e^{-\xi\frac{d}{\epsilon^2}})$  uniform in $n$ and depending only on $\dlm, \drm$. Indeed, if \eqref{xi} holds, combining with \eqref{tan} we get
\begin{align}\nonumber
\mathbb{E}_{\text{LDPC},\noise}[\langle \sigma_o\rangle_{P}]
=&\bE_{\text{LDPC},l}[\tanh(l+\Delta^{(d)})] +O(\frac{\gamma^d}{n}) 
\\ \nonumber &
+ O(e^{-\xi\frac{d}{\epsilon^2}})
\end{align}
and  the theorem  follows by taking first the limit $n\to+\infty$ and then $d\to+\infty$.

Formula \eqref{xi} follows directly from the next two lemmas.
Let $C_d(o)$ denote the circle of variable nodes at distance $=d$ from $o$.
Call $\langle -\rangle_{N_d(o)}^{+}$ the LDPC Gibbs measure associated to the graph 
$N_d(o)$  with $\sigma_j=+1$ "boundary condition" for  $j\in C_d(o)$. First we will show
\vskip 0.25cm
\begin{lemma}[Cutting a piece of the Tanner graph]\label{lemma1}
For $\epsilon^2\leq \epsilon_1^2k^{-2}(\ln k)^{-1}$ 
\begin{equation}\nonumber
\mathbb{E}_{\noise}[\langle \sigma_o\rangle_{P}]=\mathbb{E}_{\noise}[\langle \sigma_o\rangle_{N_d(o)}^+] + O(e^{-\xi \frac{d}{\epsilon^2}})
\end{equation}
where $\xi >0$ and $O(e^{-\xi \frac{d}{\epsilon^2}})$ depend only on $\dlm,\drm$. In particular they are independent of $n$.
\end{lemma}
\vskip 0.25cm
The second step is to show that for $\epsilon$ small enough the soft estimate of the bit at $o$ is independent from boundary conditions.
\vskip 0.25cm
\begin{lemma}[Independence from Boundary Conditions]\label{lemma2}
Under the same conditions than in lemma \ref{lemma1}
\begin{equation}\nonumber
\mathbb{E}_{\noise}[\langle \sigma_o\rangle_{N_d(o)}]=\mathbb{E}_{\noise}[\langle \sigma_o\rangle_{N_d(o)}^+] + O(e^{-\xi\frac{d}{\epsilon^2}})
\end{equation}
\end{lemma}
\vskip 0.25cm
\noindent{\it Proof of Lemma \ref{lemma1}}. We first introduce new interpolating Gibbs measures.
Label the variable nodes in $C_d(o)$ in some arbitrary order $C_d(o)=\{1,2,...,N\}$ and assume these bits are transmitted through a BIAWGN channel with noise vector $\nu^N=(\nu_1,...,\nu_N)$ with $0\leq\nu_k\leq \epsilon$ (here $\nu_k^2$ is the noise variance). Set $\widehat\nu^j=(0,...,0, \nu_j,\epsilon,...,\epsilon)$ for $j=1,...,N$. The interpolating Gibbs measures $\langle -\rangle_P^{\widehat\nu^j}$ are defined on the full Tanner graph with noise vectors $\widehat\nu^j$ for bits in $C_d(0)$ and noise $\epsilon$ for all other bits. A crucial remark is that for $\nu^N=(0,...,0)=\underline 0$
\begin{equation}\label{remark}
\mathbb{E}_{l^n}[\langle \sigma_o\rangle_P^{\nu^N=\underline 0}] = \mathbb{E}_{l^n}[\langle \sigma_o\rangle_{N_d(o)}^+]
\end{equation}
Proceeding similarly to \cite{KuMa08itw} we apply iteratively the fundamental theorem of calculus,
\begin{align}\nonumber
\mathbb{E}_{\noise}[\langle \sigma_o\rangle_{P}]  =
\mathbb{E}_{\noise}[\langle \sigma_o\rangle_P^{\nu^N=\underline 0}] +
\sum_{j=1}^N\int_0^\epsilon d\nu_j \frac{d}{d\nu_j}
\mathbb{E}_{\noise}[\langle \sigma_o\rangle_P^{\widehat\nu^j}]
\end{align}
For the BIAWGN channel we have the remarkable formula \cite{MacTransactions072}
\begin{equation}\nonumber
\frac{d}{d(\nu_j^{-2})}
\mathbb{E}_{\noise}[\langle \sigma_o\rangle_P^{\widehat\nu^j}] =
\mathbb{E}_{\noise}\bigl[(\langle \sigma_o
\sigma_j\rangle_{P}^{\widehat\nu^j} - \langle \sigma_o \rangle_P^{\widehat\nu^j} \langle
\sigma_j\rangle_{P}^{\widehat\nu^j})^2\bigr]
\end{equation}
Then using \eqref{remark} we obtain {\it the sum rule}
\begin{align}\nonumber
\mathbb{E}_{\noise}&[\langle \sigma_o\rangle_{P}]  = \mathbb{E}_{\noise}[\langle \sigma_o\rangle_{N_d(o)}^{+}]
\\ \nonumber &
-2\sum_{j=1}^N \int_0^\epsilon \frac{d\nu_j}{\nu_j^{3}}
\mathbb{E}_{\text{LDPC},\noise}\bigl[(\langle \sigma_o
\sigma_j\rangle_{P}^{\widehat\nu^j} - \langle \sigma_o \rangle_P^{\widehat\nu^j} \langle
\sigma_j\rangle_{P}^{\widehat\nu^j})^2\bigr]
\end{align}
Now we apply the generalized form of theorem \ref{theorem1}, namely eq \eqref{general} (with possibly different numerical constants)
\begin{equation}\nonumber
\mathbb{E}_{\noise}[\langle \sigma_o
\sigma_j\rangle_{P}^{\widehat\nu^j}  -  \langle \sigma_o \rangle_P^{\widehat\nu^j} \langle
\sigma_j\rangle_{P}^{\widehat\nu^j}]
\leq 
c_1 e^{-\frac{c}{\nu_j^2}}e^{-\frac{c_2}{\epsilon^2 k}d}
\end{equation}
Note that the prefactor $e^{-\frac{c}{\nu_j^2}}$ is important in order to get convergent integrals in the sum rule.
For the number of boundary terms we have $N\leq k^d$ which leads to
 the result 
of the lemma for $\epsilon^2\leq \epsilon_1^2 k^{-2}(\ln k)^{-1}$.
\vskip 0.25cm
\noindent{\it Proof of Lemma \ref{lemma2}}. The proof is similar to that of Lemma \ref{lemma1} with $\langle-\rangle_P$ replaced by $\langle - \rangle_{N_d(o)}$.

\section{Discussion}

Consider code ensembles such that the MAP-GEXIT curve has only one discontinuity at 
$\epsilon_{\text{MAP}}$ and vanishes for $\epsilon<\epsilon_{\text{MAP}}$. Because of the perturbative nature of the cluster expansion our estimates for theorem \ref{theorem1} only work much below 
$\epsilon_{\text{MAP}}$. What is the exact range of validity for the decay of the theorem is an open question. Let now $\epsilon_{\text{BP}}$ be the Belief Propagation threshold.
We know that theorem \ref{theorem2} cannot be valid for 
$\epsilon_{\text{BP}}<\epsilon<\epsilon_{\text{MAP}}$ since in this range the BP and MAP estimates differ. In view of the sum rule in the proof of Lemma \ref{lemma1} this means that for this range
the decay of correlations (even if exponential) cannot overcome the exponential growth of the number of nodes in $C_d(o)$. An interesting question is to determine if the smallest $\epsilon_*$ for which this happens has a clear algorithmic significance and if it is in any way related to $\epsilon_{BP}$.

Consider now the case of cycle codes, or of codes with sufficient fraction of degree two variable nodes (and no nodes of degree one), such that the GEXIT function is equal to zero for 
$\epsilon\leq \epsilon_{\text{MAP}}$, is non zero for 
$\epsilon\geq \epsilon_{\text{MAP}}$ while it remains continuous at $\epsilon_{\text{MAP}}$ (the curve may have a discontinuity at higher noise value $\epsilon_c$). Although in this case the statement of theorem \ref{theorem2} may be valid for some range of $\epsilon$ above $\epsilon_{\text{MAP}}$,  our proof only works only below $\epsilon_{\text{MAP}}$. 
This can be explicitly seen from  Lemma 
\ref{lemma2} and the fact $\mathbb{E}_{\noise}[\langle \sigma_o\rangle_{N_d(o)}^+\vert N_d(o)\,\text{is a tree}]=1$ which imply that our proof only works in a range were the GEXIT function vanihes. Our analysis is not powerful enough to capture any interesting behavior for the GEXIT function for $\epsilon_{\text{MAP}}<<\epsilon<\epsilon_{c}$. 

Finally, consider the case of ensembles with some fraction of degree one nodes and a GEXIT function that {\it does not vanish} all the way down to $\epsilon\to 0$ (with possibly a discontinuity at some $\epsilon_c$). An example is given by LDPC ensembles with Poisson degree distribution for variable nodes. Note that here $\mathbb{E}_{\noise}[\langle \sigma_o\rangle_{N_d(o)}^+\vert N_d(o)\,\text{is a tree}]\neq 1$ because the tree still contains leaves (at distance $< d$ from $o$)  with free boundary conditions. In this case theorem \ref{theorem2} really captures a non trivial behavior of he GEXIT curve for small $\epsilon$. It extends to other channels previous results
\cite{CyRuMon07isit}, \cite{KuKoMa} that had been obtained only for the BEC. This also proves that the replica solution is indeed correct for channels other than the BEC.

\section{Acknowledgment}

S.K acknowledges the support from the Fonds National Suisse pour la Recherche Scientifique, grant no 200020-113412. We would like to thank Cyril M\'easson for discussions on duality and Hamed Hassani for a discussion that led to a simplification of the proof of theorem \ref{theorem2}.


\begin{thebibliography}{1}

\bibitem{KuMa07TC} S. Kudekar, N. Macris, "Decay of correlations: an application to low density parity check codes", \emph{5th International Symposium on Turbo Codes and Related Topics}, pp. 13-18 (Lausanne 2008)

\bibitem{Georgii} H. O. Georgii, "Gibbs measures and phase transitions", \emph{de Gruyter Studies in Mathematics 9}  (1988).

\bibitem{KuMa08itw} S. Kudekar, N. Macris, "Proof of replica formulas in the high noise regime for communication using LDGM codes", \emph{Information Theory Workshop}, pp. 416-4120 (Porto 2008)

\bibitem{Brydges} D. C. Brydges, ``A short course on cluster expansions'', in les Houches summer school, Session XLIII, 1984 (K. Osterwalder and R. Stora, eds).

\bibitem{Berretti} A. Berretti, "Some properties of random Ising models", \emph{Journal of Statistical Physics}, vol 38 pp. 483-496 (1985)

\bibitem{Frohlich} J. Fr\"ohlich, "Mathematical aspects of disordered systems", in les Houches summer school, Session XLIII, 1984 (K. Osterwalder and R. Stora, eds).

\bibitem{BrinkAshikmin} A. Ashikmin, G. Kramer, S. ten Brink, "Extrinsic information transfer functions: model and erasure channel property" \emph{IEEE Trans. Inform. Theory}, vol 50, pp. 2657-2673 (2004) 

\bibitem{Montanari} A. Montanari, "Tight bounds for LDPC and LDGM codes under MAP decoding", \emph{IEEE Trans. Inform. Theory}, vol 51 pp. 3221-3246 (2005)

\bibitem{KuMa06isit} S. Kudekar, N. Macris, "Sharp bounds for MAP decoding of general irregular LDPC codes", \emph{ISIT} pp. 2259-2263, (Seattle 2006)

\bibitem{CyRuMon07isit} C. M\'easson, A. Montanari, R. Urbanke, "Asymptotic rate versus design rate", \emph{ISIT} pp. 1541-1545 (Nice 2007)

\bibitem{KuKoMa} S. Kudekar, S. Korada, N. Macris, "Exact solution for the conditional entropy of Poissonian LDPC codes over the binary erasure channel", \emph{ISIT} pp. 1016-1021 (Nice 2007)

\bibitem{MacTransactions07} N. Macris,
"Griffiths-Kelly-Sherman correlation inequalities: a useful tool in the theory of error correcting codes", \emph{IEEE Trans. Inform. Theory}, vol 53 pp. 664-683 (2007)

\bibitem{MacTransactions072} N. Macris, "Sharp bounds on generalized EXIT functions" \emph{IEEE Trans. Inform. Theory}, vol 53 pp. 2365-2375
(2007)

\bibitem{Forney} D. G. Forney, ``Codes on graphs: normal realizations'', \emph{IEEE Trans. Inform. Theory}, vol 47 pp. 520-548 (2001)


\end{thebibliography}
\end{document}